\documentclass[aps,prl,twocolumn,groupedaddress,showpacs,floatfix]{revtex4}
\usepackage{graphicx}
\usepackage{subfigure}
\usepackage{amsmath}
\usepackage[dvips]{color}
\def\ket#1{|#1\rangle}

\begin{document}

\title{ Two-field optical methods to control magnetic Feshbach resonances}

\author{A. Jagannathan$^{1,2}$, N. Arunkumar$^{1}$, J. A. Joseph$^{1}$, and J. E. Thomas$^{1}$}
\affiliation{$^{1}$Department of  Physics, North Carolina State University, Raleigh, NC 27695}
\affiliation{$^{2}$Department of Physics, Duke University, Durham, NC 27708}


\date{\today}

\begin{abstract}
	  Using an optically-trapped mixture of the two lowest hyperfine states of a $^6$Li Fermi gas, we observe two-field optical tuning of the narrow Feshbach resonance by up to 3 G and an increase in spontaneous lifetime near the broad resonance from $0.5$ ms to $0.4$ s. We present a new model of light-induced loss spectra, employing  continuum-dressed basis states, that agrees in shape and  magnitude with measurements for both broad and narrow resonances.
	
\end{abstract}

\maketitle

Optical control methods offer tantalizing possibilities  for creating ``designer" two-body interactions in ultra-cold atomic gases, with both high spatial resolution and high temporal resolution. By controlling the elastic scattering length, the inelastic scattering length, and the effective range, optical methods enable control of few-body and many-body systems, opening new fields of research.

Single optical field methods have been used to control Feshbach resonances, with large detunings to suppress spontaneous scattering, leading to limited tunability~\cite{WalravenOptTuning,JulienneOptTuning,PLettOFB,EnomotoOFB,TheisOFB,YamazakiSpatialMod,
RempeOptControl, PotassiumSingleoptical, ChinMagicOptControl}.  The first experiments, by Lett and collaborators in a Na Bose gas, demonstrated optically-induced Feshbach resonances, coupling the ground and excited molecular states in the input channel~\cite{PLettOFB}. Recently, Chin and coworkers observed suppression of both spontaneous scattering and the polarizability of Cs atoms, by tuning between the D1 and D2 lines~\cite{ChinMagicOptControl}. This  method suppresses unwanted optical forces and achieves a lifetime up to 100 ms with rapid but limited tuning, by modulating the intensity of the control beam.

Building on ideas suggested by Bauer et al.,~\cite{RempeOptControl} and by Thalhammer et al.,~\cite{ThalhammerTwoFieldOFB},  we are developing two-field optical control methods~\cite{WuOptControl1,WuOptControl2} that create a molecular dark state in the closed channel of a magnetic Feshbach resonance, as studied recently in dark state spectroscopy~\cite{MadisonDarkState}. This approach is closely related to electromagnetically induced transparency (EIT)~\cite{EITHarris}, where quantum interference suppresses unwanted optical scattering. In contrast to single field methods, our two-field methods are applicable to broad Feshbach resonances, as they produce relatively large tunings for a given loss rate. Further, the two-field methods  produce narrow energy-dependent features in the scattering phase shift, enabling control of the effective range~\cite{WuOptControl2}. Analogous to the EIT method of enhancing optical dispersion in gases with suppressed absorption~\cite{EITHarris}, the effective range can be modified in regions of highly suppressed optical scattering.

Implementation of optical control methods requires an understanding of the optically-induced level structure and energy shifts, which depend on the relative momentum of a colliding atom pair. Our original theoretical approach~\cite{WuOptControl1,WuOptControl2} and that of other groups~\cite{RempeOptControl} employed adiabatic elimination of an excited molecular state amplitude, which fails for very broad resonances  where the hyperfine coupling constant is large. This unresolved issue has been noted previously~\cite{Bauer2009}.

\begin{figure}[htb]
\centering
\includegraphics[height = 2.10 in]{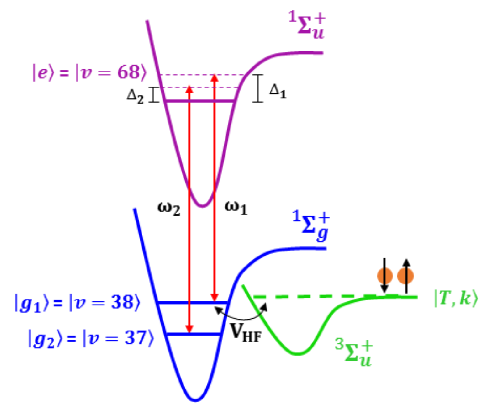}
\caption{Basic level scheme for the two-field optical technique. Optical fields of frequencies $\omega_1$ (detuning $\Delta_1$) and $\omega_2$ (detuning $\Delta_2$) , respectively, couple two singlet ground molecular states $\ket{g_1}$  and $\ket{g_2}$  to the singlet excited molecular state $\ket{e}$; $V_{HF}$ is the hyperfine coupling between the incoming atomic pair state in the open triplet channel $\ket{T,k}$ and $\ket{g_1}$, which is responsible for a magnetically controlled Feshbach resonance. \label{fig:level}}
\end{figure}

In this Letter, we demonstrate large shifts of magnetic Feshbach resonances and strong suppression of spontaneous scattering in measurements of two-field light-induced loss spectra. Further, we present a new theoretical approach to describe control of broad and narrow Feshbach resonances  in a unified manner, replacing a ``bare" state description by a more natural description in terms of ``continuum-dressed" states that incorporate the hyperfine coupling into the basis states. Using the measured Rabi frequencies, the predicted relative-momentum averaged loss spectra agree in shape and  magnitude with data for both broad and narrow resonances.

The basic level scheme for the two-field optical technique is shown in Fig.~\ref{fig:level}. An optical field with Rabi frequency $\Omega_1$ and  frequency $\omega_1$ couples the ground vibrational state $\ket{g_1}$ of the $^1\Sigma^+_g\,$ potential to the excited vibrational state $\ket{e}$ of the $^1\Sigma^+_u\,$ potential. A second optical field with Rabi frequency $\Omega_2$ and  frequency $\omega_2$ couples a lower lying ground vibrational state $\ket{g_2}$ to the excited vibrational state $\ket{e}$. The $\omega_1$ beam results in a light shift of state $\ket{g_1}$ as well as atom loss due to photoassociation from the triplet continuum $\ket{T,k}$ to the excited state $\ket{e}$. The $\omega_2$ beam suppresses atom loss through destructive quantum interference. In a magnetic field $B$, the triplet continuum $\ket{T,k}$  tunes downward $\propto$ $2\mu_B\,B$, where $\mu_B\,$ is the Bohr magneton, $\mu_B/h\simeq h\times 1.4$ MHz/G. $\ket{T,k}$ and $\ket{g_1}$ have a hyperfine coupling constant $V_{HF}$, producing a Feshbach resonance. For our experiments with $^6$Li,  $\ket{g_1}$ and $\ket{g_2}$ are the $v = 38$ and $v = 37$ ground vibrational states and $\ket{e}$ is the $v = 68$ excited vibrational state, which decays at a rate $\gamma_e=2\pi\times 11.8$ MHz.

The detunings of the optical fields that couple state $\ket{g_1}$ and $\ket{g_2}$ to $\ket{e}$ are $\Delta_1$ and $\Delta_2$, respectively. The single photon detuning of the $\omega_1$ beam for the $\ket{T}\rightarrow\ket{e}$ transition is a function of magnetic field and can be defined at a reference magnetic field $B_{ref}$ as $\Delta_e = \Delta_L-2\mu_B(B-B_{ref})/\hbar$, where $\Delta_L$ is the detuning of the optical field when $B=B_{ref}$. The two-photon detuning for the $\ket{T}-\ket{e}-\ket{g_2}$ system is $\delta = \Delta_e - \Delta_2$.

\begin{figure}[htb]
	\centering
		\includegraphics[width = 3.2 in]{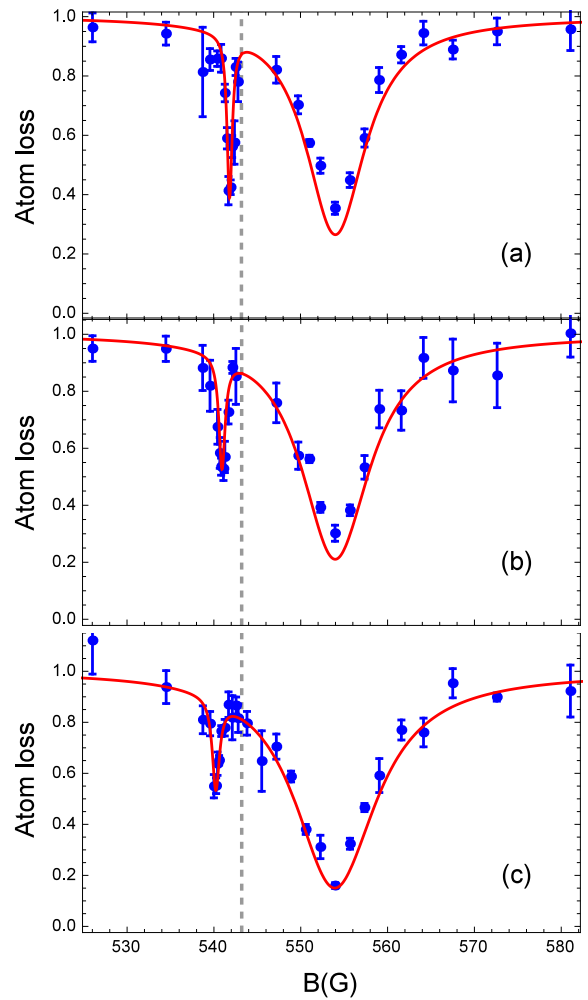}\\
		\caption{Shifting the narrow Feshbach resonance at $543.2$ G using a single optical field. The large loss peak on the right arises from the broad resonance; The left loss peak arises from the shifted narrow resonance. Vertical dashed line: Position of the unshifted narrow resonance. The background (non-optical) three-body loss near 543.2 G has been suppressed for clarity. Pulse duration $\tau =5.0$ ms; $T=5.4 \,\mu$K; $\Delta_L =\Delta_1 = 30.2$ MHz; (a)  $\Omega_1=2.00\,\gamma_e$; $B^\prime_{res} = 541.9 $ G (b)  $\Omega_1=2.65\,\gamma_e$; $B^\prime_{res} = 541.0$ G (c) $\Omega_1=3.10\,\gamma_e$; $B^\prime_{res} = 540.2$ G; $\gamma_e=2\pi\times 11.8$ MHz. Blue dots: Experiment; Solid red curves: Continuum-dressed state model~\cite{SupportOnline}. \label{fig:Narrow540}}
		\end{figure}

\begin{figure*}[htb]
		\centering
			\includegraphics[width=7in]{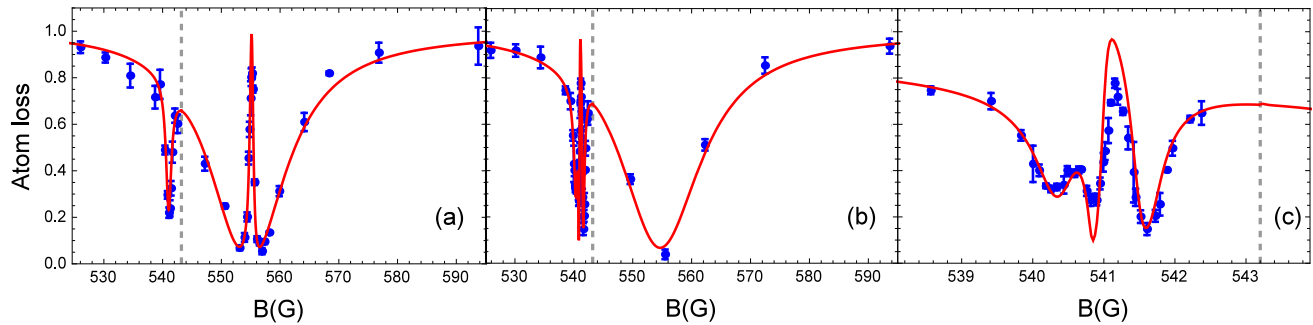}\\
			\caption{Loss suppression using the two-field optical technique. The $\omega_1$ beam shifts the narrow Feshbach resonance and the frequency of the $\omega_2$ beam is chosen to suppress loss from (a) the broad resonance (b) the shifted narrow Feshbach resonance; (c) shows an expanded view of (b) near the loss suppression region. Pulse duration $\tau =5.0$ ms;  $T=4.5\,\mu$K; $\Omega_1=2.6\,\gamma_e$; $\Omega_2=0.8\,\gamma_e$; $\Delta_L = \Delta_1 = 30.2$ MHz; $B^\prime_{res} = 541.1$ G. Solid red curves: Continuum-dressed state model~\cite{SupportOnline}.\label{fig:NarrowEIT}}
\end{figure*}
We prepare a 50:50 mixture of $^6$Li atoms in the two lowest hyperfine levels, $\ket {1}$ and $\ket {2} $ in a CO$_2$ laser trap with trap frequencies ($\omega_x, \omega_y, \omega_z) = 2\pi\times(3100,3350,120)$ Hz. The $\ket {1}$-$\ket {2}$ mixture of $^6$Li has a broad Feshbach resonance at $B_\infty=832.2$ G~\cite{JochimPreciseFeshbach,BartensteinFeshbach} of width $\Delta B = 300$ G due to strong hyperfine coupling of the triplet continuum to the ``broad" singlet state $\ket{g_1}$$_{B}$~\cite{WuOptControl2}. In addition, there is a narrow Feshbach resonance at $543.2$ G of width $\Delta$B = 0.1 G~\cite{OHaraNarrowFB} due to  weak second order hyperfine coupling of the triplet continuum $\ket{T}$ to the ``narrow" singlet state $\ket{g_1}$$_{N}$~\cite{WuOptControl2}. After forced evaporation, and re-raising the trap to full trap depth, we have approximately $10^5$ atoms per spin state for our experiments. We generate both the $\omega_1$ and $\omega_2$ beams from diode lasers, locked to a stabilized cavity near 673.2 nm. The relative frequency is $\simeq 57$ GHz~\cite{MadisonDarkState} with a jitter  $<50$ kHz. The absolute frequency stability is $<$100 kHz. Both laser beams are sent through fibers and focused onto the optically-trapped atom cloud with both beams polarized along the bias magnetic field $z$-axis~\cite{SupportOnline}.

Initially, we use a single optical field ($\omega_1$ beam) to observe the shift of the narrow Feshbach resonance in the atom loss spectra. After forced evaporation at 300 G, the magnetic field is swept to the field of interest and allowed to stabilize. Then the $\omega_1$ field is shined on the atoms for 5 ms, with a detuning $\Delta_L = 30.2$ MHz (with $\Delta_L \equiv 0$ for $B = B_{ref} = 543.2$ G), after which the atoms are imaged at the field of interest to determine the density profile and the atom number.

Single field atom loss spectra versus magnetic field,  Fig.~\ref{fig:Narrow540}, exhibit two loss peaks: i) A broad peak  arises at 554 G, where the $\omega_1$ optical field is resonant ($\Delta_e = 0$)  with the $\ket{T}\rightarrow\ket{e}$ transition. Here, the transition  arises from the hyperfine coupling of $\ket{T}$ to $\ket{g_1}$$_{B}$, far from the resonance at 832.2 G. ii) A narrow peak below $543.2$ G occurs as the magnetic field tunes the triplet continuum near $\ket{g_1}_{N}$, which is light-shifted in energy (to $B'_{res}$) due to the $\omega_1$ optical field, detuned from the $\ket{g_1}_N\rightarrow \ket{e}$ transition by $\Delta_1 = \Delta_L = 30.2$ MHz (see Fig.~\ref{fig:level}). In this case, the transition strength is resonant, while the $\omega_1$ optical field is off-resonant with the $\ket{T}\rightarrow\ket{e}$ transition by $\Delta_L = 30.2$ MHz. For a $\Omega_1 = 3.1\gamma_e$, the narrow  resonance is shifted downward by 3.0 G, approximately 30 times the width.  The continuum-dressed state model (solid red line), simultaneously reproduces the shift of the narrow resonances and the amplitudes of both the narrow and broad resonances, using the measured $\Omega_1$~\cite{SupportOnline}.

For the two-field loss-induced suppression measurements, Fig.~\ref{fig:NarrowEIT}, the $\omega_2$ optical field is applied. After sweeping to the magnetic field of interest, the $\omega_2$ beam of intensity of 0.4 kW/cm$^{2}$ is adiabatically turned on over 30 ms.  The $\omega_2$ beam creates an optical dipole trap and provides additional confinement in the z-direction, due to its high intensity. This changes the axial trap frequency from 120 Hz to 218 Hz, with negligible change in the radial trap frequencies. The $\omega_1$ beam is then turned on for 5 ms, after which both beams are turned off abruptly.  The detuning $\Delta_2$ of the $\omega_2$ beam can be chosen to suppress loss either at the broad peak or the narrow peak. For $\Delta_2 = \Delta_e =0$, the loss is suppressed at the center of the broad peak Fig.~\ref{fig:NarrowEIT}a. For $\Delta_2 = \Delta_1 + 2\mu_B\,(543.2-B^\prime_{res})$, the loss is suppressed at the center of the shifted narrow peak, Fig.~\ref{fig:NarrowEIT}b and~\ref{fig:NarrowEIT}c. The  loss spectra clearly demonstrate the Feshbach resonance can be strongly shifted and that loss can be strongly suppressed using  two-field methods. The continuum-dressed state model (red solid curves) predicts the features for all three data sets using the same Rabi frequencies $\Omega_1$ and $\Omega_2$, which are close to the predicted values~\cite{SupportOnline}. We note that the predicted central peaks in Fig.~\ref{fig:NarrowEIT}c are somewhat larger than the measured values, which may arise from  jitter in the two-photon detuning and  intensity variation of the $\Omega_2$ beam across the atom cloud.

\begin{figure}[htb]
		\centering
		\includegraphics[width=3.0in]{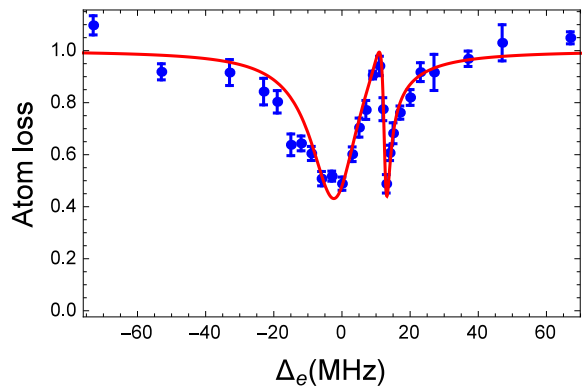}
		\caption{Loss suppression near the broad resonance at $832.2$ G, for  $B=840$ G, as a function of single photon detuning by sweeping the $\omega_1$ laser frequency. Pulse duration $\tau =5.0$ ms; $T=14.8\,\mu$K $\Omega_1=1.36\,\gamma_e$; $\Omega_2=0.9\,\gamma_e$; $\Delta_2=10.0$ MHz. Maximum suppression occurs for $\Delta_e = \Delta_2 =10.0$ MHz, where $\delta_e=0$. Solid red curve: Continuum-dressed state model~\cite{SupportOnline}. \label{fig:Broad840FreqScan}}
\end{figure}

\begin{figure}[htb]
		\centering
		\includegraphics[width=3.0in]{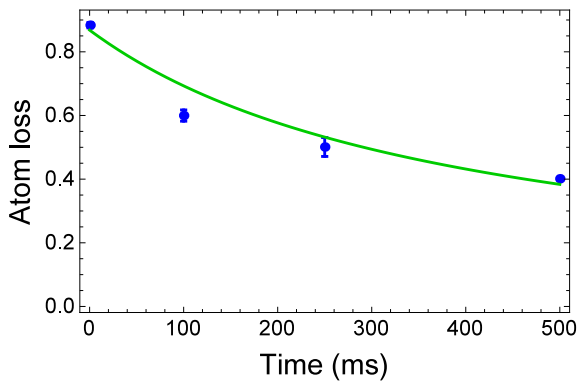}
		\vspace{0.15in}\caption{Number of atoms in state $|1\rangle$ versus time with $\Omega_1\ = 0.65\,\gamma_e$ and $\omega_1$ tuned to cause loss at 841 G.  $\Omega_2=0.9\,\gamma_e$ and $\omega_2$ is tuned to suppress loss near 841 G. With $\Omega_2=0$, the corresponding decay time is $\simeq 0.5$ ms.  $\gamma_e=2\pi\times 11.8$ MHz is the radiative decay rate; $T=4.5\,\mu$K.  Solid green curve: $N(t)=N(0)/(1+\gamma t)$, where $\gamma =2.5\,s^{-1}$. \label{fig:Broad840TimeScan}}
\end{figure}

We have also measured light-induced loss and loss suppression as a function of the $\omega_1$ laser frequency near the broad resonance. Data for $B=840$ G is shown in Fig.~\ref{fig:Broad840FreqScan}. When the $\omega_1$ beam is tuned to achieve two-photon resonance, the loss is highly suppressed. For this experiment, $\Delta_2=10$ MHz. Since the maximum suppression occurs for $\Delta_e = \Delta_2$,  we observe an asymmetric EIT window. To examine the loss suppression further, we measure the number of atoms as a function of time with the magnetic field tuned to the suppression point.  We observe dramatic suppression of loss using the two-field method, achieving an increase of the inelastic lifetime near the broad resonance of $^6$Li from $0.5$ ms with a single laser field to $400$ ms with the two-field method, Fig.~\ref{fig:Broad840TimeScan}, limited by jitter in the two-photon detuning.\\

Measured light-induced loss spectra are compared to predictions by calculating the atom loss rate~\cite{SupportOnline}. For a 50-50 mixture of two hyperfine states,
the total density decays according to $\dot{n}({\mathbf{r}},t)=-\frac{1}{2}\langle K_2(k,{\mathbf{r}})\rangle\,[n({\mathbf{r}},t)]^2$.
Here, the angle brackets in $\langle K_2(k,{\mathbf{r}})\rangle$ denote an average over the relative momentum $\hbar k$ distribution. As the Rabi frequencies that determine $K_2$ generally vary in space, we include an additional position dependence in $\langle K_2\rangle$.  For simplicity, we assume in this paper a classical Boltzmann distribution, which is applicable in the high temperature regime employed in the measurements and defer treatment of quantum degeneracy and many-body effects to future work.

\begin{figure}[htb]
		\centering
		\includegraphics[height = 1.3in]{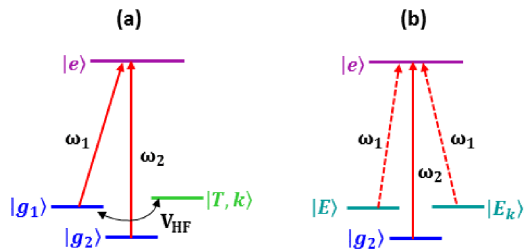}
		\caption{Level schemes in (a) ``bare-state" and (b) ``continuum-dressed-state" bases. $\ket{g_1}$, $\ket{g_2}$, and $\ket{e}$ are the bare molecular states in the energetically closed (singlet) channel. $\ket{T,k}$ is a bare continuum state in the open (triplet) channel. The hyperfine interaction $V_{HF}$ couples the bare molecular state $\ket{g_1}$  and the continuum states $\ket{T,k}$, creating the ``continuum-dressed" bound state $\ket{E}$ and the (Feshbach resonance) scattering state $\ket{E_k}$. \label{fig:basis}}
\end{figure}

The loss rate constant $K_2(k)$ is calculated from the optically-modified scattering state in the continuum-dressed state basis~\cite{SupportOnline}. In previous calculations~\cite{WuOptControl1,WuOptControl2,RempeOptControl}, interaction of the colliding atom pair with the optical fields is described in the ``bare" state basis, Fig.~\ref{fig:basis}a, with singlet states, $\ket{g_1}$, $\ket{g_2}$, and $\ket{e}$, and triplet continuum $\ket{T,k}$. Using the continuum-dressed state basis, Fig.~\ref{fig:basis}b, the bare states $\ket{g_1}$ and $\ket{T,k}$, are replaced by  the dressed bound state $\ket{E}$ and the Feshbach resonance scattering state $\ket{E_k}$. These dressed states already contain the hyperfine coupling constant $V_{HF}$,  permitting consistent adiabatic elimination of the excited state amplitude $\ket{e}$, even for broad Feshbach resonances where $V_{HF}$ is large. From the scattering state, we determine the corresponding two-body scattering amplitude $f(k)$, which yields $K_2(k)$ from the inelastic cross section. The new model shows that the light-shifts arising from the $\Omega_1$ beam have a different relative momentum ($k$) dependence for  broad resonances than for narrow resonances. Further, it reproduces  previous calculations~\cite{WuOptControl1,WuOptControl2,RempeOptControl} that are valid only for narrow resonances and avoids  predictions of a spurious broad resonance at $B_\infty$ that arises when narrow resonance results are incorrectly applied to broad resonances.

In conclusion, we have demonstrated that two-optical field methods can produce large shifts of magnetic Feshbach resonances and strong suppression of spontaneous scattering, enabling flexible control of the tradeoff between loss and tunability.  We have established a continuum-dressed state model that fits  our measured loss spectra in shape and  magnitude, using the measured Rabi frequencies and trap parameters. A key result of this model is the prediction of the relative momentum dependence of the light-induced level shifts for both broad and narrow resonances, resolving a long-standing issue with predictions for broad resonances. Using the predicted relative momentum dependence of the scattering amplitude, the model can be used to examine optical control not only of the scattering length, but of the effective range, which will be experimentally studied in future work.


\appendix
\section{Supplemental Material}
\subsection{Theory}
\label{sec:theory}

To compare the measured loss spectra with predictions, we find the two-field optically-induced change in the two-body scattering properties of an ultra-cold atomic gas near a magnetic Feshbach resonance.
The scattering length and effective range are determined by the relative-momentum-dependent two-body scattering amplitude,
\begin{equation}
f=\frac{e^{2i\delta_s(k)}-1}{2ik}=\frac{|a_{bg}|}{x\cot\delta_s(x)-ix},
\label{eq:scattampl}
\end{equation}
where $\hbar k$ is the relative momentum and the total scattering phase shift $\delta_s(k)=\Delta(k)+\Phi(k)$ is  found from
\begin{eqnarray}
x\cot\delta_s(x)&=&\frac{x\cot\Delta(x)\,x\cot\Phi(x)-x^2}{x\cot\Phi(x)+x\cot\Delta(x)}\nonumber\\
&\equiv& q'(x)+iq''(x).
\label{eq:q}
\end{eqnarray}
In Eqs.~\ref{eq:scattampl} and~\ref{eq:q}, we have defined the dimensionless momentum $x=k|a_{bg}|$, with $a_{bg}<0$ the background scattering length. $\Delta(x)$ arises from the background Feshbach resonance, (Eq.~\ref{eq:38.1} below)  while $\Phi(x)$  arises from the optical fields (Eq.~\ref{eq:38.4} below). For later use, we have defined the real and imaginary parts of $x\cot\delta_s(x)$, $q'(x)$ and $q''(x)$, which determine all of the optically-controlled scattering parameters.

The phase shifts $\Delta$ and $\Phi$ are determined from the scattering state in the continuum dressed state basis,
comprising the dressed bound state $\ket{E}$, the Feshbach resonance continuum state $\ket{E_k}$, and the singlet states $\ket{g_2}$ and $\ket{e}$, as shown Fig.~6b of the main text. This method enables adiabatic elimination of the excited electronic state $\ket{e}$ amplitude, even for broad Feshbach resonances where the hyperfine coupling constant is large, as noted above, provided that the Rabi frequencies are not too large. From the asymptotic ($r\rightarrow\infty$) scattering state, we determine the  scattering phase shift $\delta_s(k)$ using a method analogous to Ref.~\cite{WuOptControl2} with $\langle r|E_k\rangle\rightarrow\sin [kr+\Delta(k)]/(kr\sqrt{V})$ playing the role of the background triplet continuum states $\langle r|k\rangle$.

In the absence of optical fields ($\Omega_1\rightarrow 0$), the model presented in the Appendix of Ref.~\cite{WuOptControl2} determines the Feshbach resonance continuum states with $E_k=E_T+\hbar k^2/m\geq E_T$, yielding
\begin{equation}
x\cot\Delta(x)=\frac{\tilde{\Delta}_0-\epsilon\,x^2}{1+\tilde{\Delta}_0-\epsilon\,x^2}.
\label{eq:38.1}
\end{equation}
where $\tilde{\Delta}_0\equiv(B-B_\infty)/\Delta B$ is the detuning for the magnetic Feshbach resonance and
$\epsilon\equiv E_{bg}/(2\mu_B\Delta B)$, with $E_{bg}=\hbar^2/(ma_{bg}^2)$, $m$ the atom mass, and $\mu_B$ the Bohr magneton. With $\Omega_1$ defined as the Rabi frequency for the singlet $\ket{g_1}\rightarrow\ket{e}$ transition in Fig.~6a of the main text, the Rabi frequency for an optical transition from $\ket{E_k}$ to $\ket{e}$ in Fig.~6b of the main text is $ \Omega_1\,|\langle g_1|E_k\rangle|$, where
\begin{equation}
|\langle g_1|E_k\rangle|^2=\frac{\epsilon|a_{bg}|^3}{2\pi^2}\,L(\tilde{\Delta}_0,x),
\label{eq:38.2}
\end{equation}
with
\begin{equation}
L(\tilde{\Delta}_0,x)=\frac{1}{(\Delta_0-\epsilon\, x^2)+x^2(1+\Delta_0-\epsilon\, x^2)^2}.
\label{eq:38.3}
\end{equation}

Using the same model, we determine the dressed bound state $\ket{E}$  by solving the Schr\"{o}dinger equation for  states of energy $E<E_T$, i.e., for $E=E_T-E_m\,E_{bg}$, where $E_m$ is the dimer binding energy in units of $E_{bg}$. In this case, the Rabi frequency for an optical transition from $\ket{E}$ to $\ket{e}$ in Fig.~6b is $\Omega_1|\langle g_1|E\rangle|$, where
\begin{equation}
|\langle g_1|E\rangle|^2=\left(1+\frac{1}{2\epsilon}\frac{1}{\sqrt{E_m}}\frac{1}{(1+\sqrt{E_m})^2}\right)^{-1}.
\label{eq:z}
\end{equation}
The dimensionless binding energy $E_m$  satisfies,
\begin{equation}
\epsilon E_m+\tilde{\Delta}_0+\frac{\sqrt{E_m}}{1+\sqrt{E_m}}=0,
\label{eq:binding}
\end{equation}
with $\epsilon$ and $\tilde{\Delta}_0$ as defined above. Here, $\tilde{\Delta}_0<0$ is required for the bound state $\ket{E}$ of binding energy $E_m>0$ to exist. Eq.~\ref{eq:z} and Eq.~\ref{eq:binding} are equivalent to the results of Ref.~\cite{StoofDressedBound}. One can verify by numerical integration that $|\langle g_1|E\rangle|^2 +\int_0^\infty 4\pi k^2 dk\,|\langle g_1|E_k\rangle|^2=1$ as it should.

Including the interaction with both optical fields, we find the asymptotic scattering state in the continuum-dressed basis, which yields the optically-induced phase shift from
\begin{equation}
x\cot\Phi(x)=-\frac{\tilde{\delta}_e(x)+\frac{\tilde{\Omega}_1^2}{4}\frac{\hbar\gamma_e}{2\mu_B\Delta B}\,S(\tilde{\Delta}_0,x)+\frac{i}{2}}{\frac{\tilde{\Omega}_1^2}{4}\frac{\hbar\gamma_e}{2\mu_B\Delta B}\,L(\tilde{\Delta}_0,x)}.
\label{eq:38.4}
\end{equation}
Here, the dimensionless detunings are given in units of the spontaneous decay rate $\gamma_e$,
\begin{eqnarray}
&\tilde{\delta}_e(x)\equiv\tilde{\Delta}_e(x)+\frac{\tilde{\Omega}_2^2}{4\tilde{\delta}(x)}\\
&\tilde{\Delta}_e(x)\equiv\frac{2\pi\nu_1}{\gamma_e}-\frac{2\mu_B}{\hbar\gamma_e}(B-B_{ref})+\frac{E_{bg}}{\hbar\gamma_e}x^2\nonumber\\
&\tilde{\delta}(x)\equiv\frac{2\pi(\nu_2-\nu_1)}{\gamma_e}+\frac{2\mu_B}{\hbar\gamma_e}(B-B_{ref})-\frac{E_{bg}}{\hbar\gamma_e}x^2,\nonumber
\label{eq:39.5}
\end{eqnarray}
where $\nu_1=0$ corresponds to a single photon $T\rightarrow e$ resonance ($\tilde{\Delta}_e=0$) at the reference magnetic field $B_{ref}$. Similarly, $\nu_2=\nu_1$ corresponds to the two-photon resonance ($\tilde{\delta}=0$) for the $g_1\rightarrow e\rightarrow g_2$ transition. The dimensionless Rabi frequency is $\tilde{\Omega}_1\equiv\Omega_1/\gamma_e$ for the $g_1\rightarrow e$ transition  and  $\tilde{\Omega}_2\equiv\Omega_2/\gamma_e$ for the $g_2\rightarrow e$ transition.

The numerator of  Eq.~\ref{eq:38.4} contains an $\Omega_1$-dependent energy shift, with
\begin{eqnarray}
\frac{S(\tilde{\Delta}_0,x)}{2\mu_B\Delta B}&\equiv& {\cal P}\int dk'\,4\pi k^{'2} \frac{|\langle g_1|E_{k'}\rangle|^2}{E_{k'}-E_k}\nonumber\\
& &-\frac{|\langle g_1|E\rangle|^2}{E_{bg}\,E_m+\frac{\hbar^2k^2}{m}}\,\Theta[B_\infty-B].
\label{eq:shift}
\end{eqnarray}
The integral term arises from the dressed continuum states, where $E_k'-E_k=\hbar^2(k'^2-k^2)/m$,  ${\cal P}$ denotes a principal part ($k'\neq k$), and $|\langle g_1|E_{k'}\rangle|^2$ is given by Eq.~\ref{eq:38.2}. The second term arises from the dressed bound state, which exists for $B<B_\infty$. The binding energy $E_T-E=E_{bg}E_m$ is determined in units of $E_{bg}$ using Eq.~\ref{eq:binding} and $\langle g_1|E\rangle|^2$ is given by Eq.~\ref{eq:z}.

Eq.~\ref{eq:shift} is reasonably complicated. However, for the broad $1-2$ resonance in $^6$Li, where $\epsilon = E_{bg}/(2\mu_B\Delta B)=0.00036<<1$, and for the  narrow $1-2$ resonance, where $\epsilon=556>>1$, we find that the dimensionless shift functions can be simplified,
\begin{eqnarray}
S(\tilde{\Delta}_0,x)&=&\frac{\tilde{\Delta}_0+(1+\tilde{\Delta}_0)x^2}
{\tilde{\Delta}_0^2+(1+\tilde{\Delta}_0)^2x^2}\quad\text{for}\,\,\epsilon <<1\nonumber\\
S(\tilde{\Delta}_0,x)&=&\frac{1}{\tilde{\Delta}_0-\epsilon\, x^2}\hspace*{0.20in}\text{for}\,\,\epsilon >>1.
\label{eq:shiftcases}
\end{eqnarray}
We confirm Eq.~\ref{eq:shiftcases} by numerical evaluation of Eq.~\ref{eq:shift} in the broad and narrow resonance limits.

The shift given by Eq.~\ref{eq:shiftcases} for the broad resonance $\epsilon <<1$ is an important new result of the continuum-dressed state approach and leads to a $k$-averaged two-body loss rate constant in agreement with our experiments.  In particular, it avoids the incorrect prediction of a fixed resonance in the momentum-integrated loss rate at $B_\infty$, which arises when the narrow resonance form of the shift is incorrectly applied to describe the broad resonance.

Now we determine the two-body loss rate constant, $K_2(k)$ from the inelastic scattering cross section,
\begin{equation}
K_2(k)=v_{rel}\,\sigma_{inel}(k)=\frac{\hbar k}{\mu}\,\sigma_{inel}(k),
\label{eq:inel}
\end{equation}
where $\mu=m/2$ is the reduced mass. For a two-component Fermi gas (where  atoms in different hyperfine states are distinguishable), the optical theorem for the total scattering cross section and the definition of the elastic scattering cross section give the inelastic scattering cross section in the form
\begin{equation}
 \sigma_{inel}=\sigma_{tot}-\sigma_{el}=\frac{4\pi}{k}\,Im[f(k)]-4\pi|f(k)|^2.
 \label{eq:inel2}
 \end{equation}
Using Eq.~\ref{eq:q} in Eq.~\ref{eq:scattampl}, the scattering amplitude is $f(x)=|a_{bg}|/[q'(x)-i(x-q''(x))]$, and
\begin{equation}
K_2(x)=\frac{8\pi\hbar|a_{bg}|}{m}\,\frac{-q''(x)}{[q'(x)]^2+[x-q''(x)]^2}.
\label{eq:rateconstant}
\end{equation}

The atom loss rate depends on the relative-momentum-averaged loss rate constant. For simplicity, we assume a classical Boltzmann distribution of relative momentum, which is applicable for the temperatures used in the experiments that are reported in this paper. We defer treatment of degeneracy and many-body effects to future work. For a classical gas at temperature $T$, the relative momentum-averaged loss rate constant is then,
\begin{equation}
\langle K_2\rangle=\frac{4}{\sqrt{\pi}}\int_0^\infty du\, u^2 e^{-u^2}K_2(u\, x_0),
\end{equation}
where $x_0\equiv k_0|a_{bg}|$ with $\hbar k_0 = \sqrt{m k_BT}$ the thermal relative momentum.

We measure the net atom loss rate for a 50-50 mixture of two hyperfine states, denoted $1$ and $2$. In this case, the densities $n_1({\mathbf{r}})=n_2({\mathbf{r}})=n({\mathbf{r}})/2$, where $n({\mathbf{r}})$ is the total density. Then, the local loss rate is
\begin{equation}
\dot{n}({\mathbf{r}},t)=-\frac{1}{2}\langle K_2({\mathbf{r}})\rangle\,[n({\mathbf{r}},t)]^2.
\label{eq:densitylossrate}
\end{equation}
As the Rabi frequencies that determine $K_2$ generally vary in space, we include an additional position dependence in $\langle K_2\rangle$. The total loss rate constant is the sum of the independent broad and narrow contributions, $K_2=K_{2B}+K_{2N}$, since the singlet states $\ket{g_1}_B$ and $\ket{g_1}_N$ that cause the broad and narrow Feshbach resonances comprise different combinations of total electron and nuclear-spin states, which are  not optically coupled~\cite{WuOptControl2}.

Eq.~\ref{eq:densitylossrate} gives the position-dependent loss rate, which is largest where the density is largest. To determine the total number of atoms $N(t)$ at a time $t$ after the  optical fields are turned on,  we need to consider possible time-dependent changes in the spatial profile of the atom cloud, which is initially gaussian in all three directions. For moderate atom loss, when the duration of the optical pulse is long compared to the oscillation frequencies in all three directions, the gaussian initial shape of the atomic distribution will be approximately maintained as the total number decreases. Eq.~\ref{eq:densitylossrate} can be integrated over all three directions to obtain $\dot{N}$. Assuming that the diameter of the atom cloud is small compared to the diameters of the optical beams, so that the Rabi frequencies are nearly independent of ${\mathbf{r}}$, we have $\dot{N}=-\Gamma\,N^2$, where
$\Gamma = \frac{\bar{n}}{2N}\langle K_2(k)\rangle$, with $\bar{n}$ the mean density. For a gaussian atomic spatial profile, where the $1/e$ widths are $\sigma_x,\sigma_y,\sigma_z$,  $\bar{n}/N=1/[(2\pi)^{3/2}\sigma_x\sigma_y\sigma_z]$, which is independent of the atom number. In this case,
\begin{equation}
N(t)=\frac{N(0)}{1+N(0)\,\Gamma t}.
\label{eq:N}
\end{equation}
For simplicity, we use Eq.~\ref{eq:N} to predict all of the spectra shown as the red curves in Figs.~2,~Figs.~3~and~4 of the main text. Although the period for motion in the z-direction is comparable to the optical pulse duration, we find that this approximation yields spectra that are in reasonably good agreement with all of the data.

\subsection{Experiment}
\label{sec:experiment}

The method for generating the optical fields for the two-field optical technique is shown in Fig.~\ref{fig:system}. The frequency of the reference laser $L_R$ is stabilized by a PDH lock to the FP cavity. The FP cavity is locked to the error voltage generated from the iodine saturation absorption signal, using a beam from the reference laser $L_R$.
\begin{widetext}
\begin{center}
\begin{figure}[htb]
\includegraphics[width=6.0in]{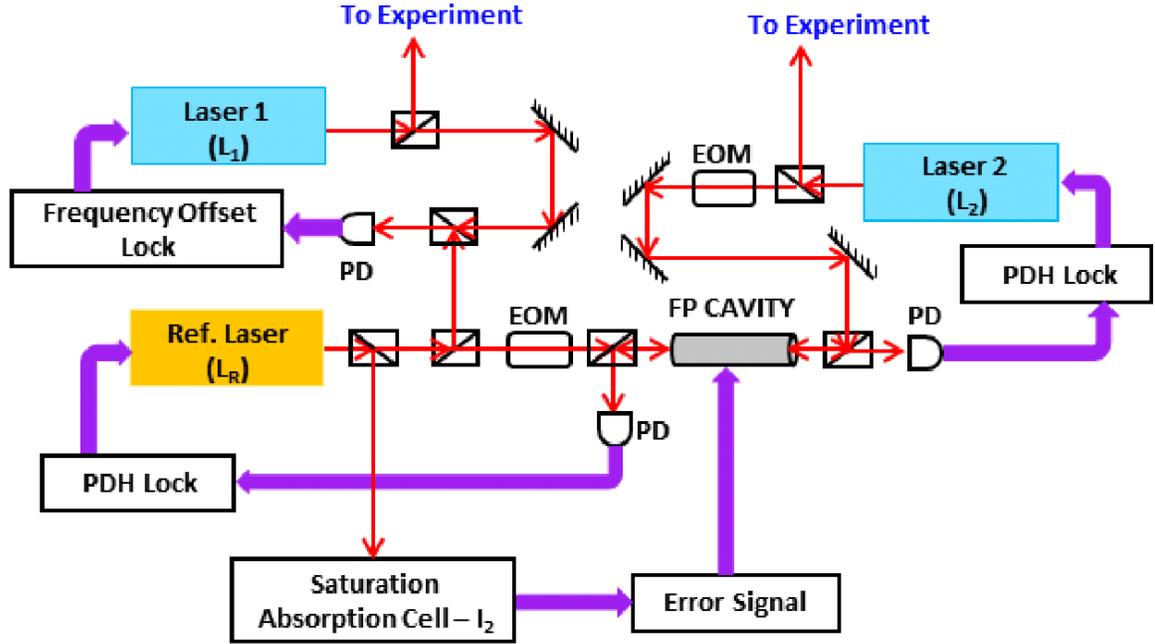}
		\caption{Diode laser system for two-field optical control. $L_R$ provides the frequency reference for the diode lasers $L_1$ and $L_2$ that generate the optical frequencies $\omega_1$ and $\omega_2$, near 673.2 nm, which excite molecular transitions from the $^6$Li singlet molecular vibrational ground states $v=38$ and $v=37$, respectively. \label{fig:system}}
\end{figure}
\end{center}
\end{widetext}
The main advantage of this setup is that it exploits the high bandwidth lock of the PDH technique to minimize the fast jitter of the diode lasers  and simultaneously provides an absolute frequency reference. Laser $L_1$ with frequency $\omega_1$ is frequency offset locked to the reference laser and generates the optical field for the $v = 38$ to $v = 68$ transition. Laser $L_2$ with frequency $\omega_2$ is locked to a different mode of the cavity and generates the optical field for the $v = 37$ to $v' = 68$ transition which is approximately 57 GHz higher in frequency from the $v = 38$ to $v' = 68$ transition. The relative frequency jitter between the lasers is  $\Delta\nu<50$ kHz and the absolute frequency stability is $<$100 kHz. As the optical linewidth of the molecular transitions is $\gamma_e=2\pi\times 11.8$ MHz, the absolute stability is not as critical as the relative stability, which limits the effective linewidth of the ground state coherence created by the two-field method.

\subsection{Determination of the trap frequencies}
\label{subsec:Calculation of trap frequency}
For the CO$_2$ laser trap alone, the trap frequencies for the $x$, $y$ and $z$ directions are determined by parametric resonance. However,
the intense $\omega_2$ beam creates a red-detuned optical dipole  trap and provides additional axial ($z$) confinement for the atom cloud, with negligible change in the transverse $x,y$ confinement. To determine the correct mean atom density,  it is important to measure the $\omega_z$ trap frequency  including the effect of the $\omega_2$ beam on the atoms. For the two-beam trap, we find that parametric resonance does not yield a high precision measurement of $\omega_z$. Instead, we use cloud size measurements to find  the total  trap frequency $\omega_{zt}$, arising from combined the CO$_2$ laser  and $\omega_2$ beam potentials.  The $\omega_1$ beam is turned off throughout the measurement. After forced evaporation in the CO$_2$ laser trap near $832$ G and re-raising the trap to full depth, the $\omega_2$ beam is adiabatically turned on over 30 ms. Then both the CO$_2$ laser and $\omega_2$ beams are abruptly turned off simultaneously and the atoms are imaged after as short time of flight to determine the $1/e$ size $\sigma_{zi}$ of the cloud just before release.  The same procedure is repeated again, but this time, only the $\omega_2$ beam is abruptly extinguished, so that the atoms are released from the $\omega_2$ trap into the CO$_2$ laser trap, where the axial frequency is $\omega_z<\omega_{zt}$. After a hold time of $200$ ms, the atom cloud reaches equilibrium and is imaged to determine its final axial width $\sigma_{zf}$ in the CO$_2$ laser trap alone. From $\sigma_{zi}$ and $\sigma_{zf}$, we determine $\omega_{zt}$ using energy conservation as follows.

Just before the $\omega_2$ beam is extinguished, the mean z-potential energy per atom is $\frac{1}{2}m\omega^2_{zt}\langle z^2\rangle$, where $\langle z^2\rangle=\frac{\sigma_z^2}{2}$ is the mean square cloud size.
As the pressure is isotropic, the total potential energy of the atoms taking into account all three directions is $\frac{3}{2}m\omega^2_{zt}\frac{\sigma^2_{zi}}{2}$. According to virial theorem~\cite{ThomasUniversal}, the mean internal energy of a unitary gas (near 832 G) is equal to the mean potential energy. Hence the internal energy of the gas at the time of release is also $\frac{3}{2}m\omega^2_{zt}\frac{\sigma^2_{zi}}{2}$. Just after extinguishing the $\omega_2$ beam, the potential energy of the atoms (in the CO$_2$ laser trap alone)  is  $\frac{3}{2}m\omega^2_{z}\frac{\sigma^2_{zi}}{2}$, since $\sigma_{zi}$ has not changed.
The total energy of the atoms immediately after the $\omega_2$ beam is turned off is then $E_{ti} = \frac{3}{2}m\omega^2_{zt}\frac{\sigma^2_{zi}}{2} + \frac{3}{2}m\omega^2_{z}\frac{\sigma^2_{zi}}{2}$. After reaching equilibrium, the total energy of the atoms  in the CO$_2$ laser trap alone is twice the mean potential energy, $E_{tf} = 3m\omega^2_{z}\frac{\sigma^2_{zf}}{2}$, according to the virial theorem. By conservation of energy $E_{ti} = E_{tf}$,
\begin{equation}
\omega^2_{zt} =  \omega^2_{z}\left(2\frac{\sigma^2_{zf}}{\sigma^2_{zi}} - 1\right),
\label{eq:freq}
\end{equation}
which gives $\omega_{zt}$ in terms of the CO$_2$ laser trap axial frequency, $\omega_{z}$.

The combined trap frequency $\omega_{zt}$ determines the initial temperature $T$ of the cloud in the total trapping potential (before release) using $k_BT=m\omega_{zt}^2\langle z^2\rangle=m\omega_{zt}^2\frac{\sigma_{zi}^2}{2}$. The cloud sizes in the $x$ and $y$ dimensions are then easily found from $\omega_x^2\sigma_{xi}^2=\omega_y^2\sigma_{yi}^2=\omega_{zt}^2\sigma_{zi}^2$, yielding $\bar{n}/N$.

\subsection{Determination of the Rabi frequencies}

We first estimate the Rabi frequency $\Omega_1$ from the predicted electric dipole transition matrix element. In  $^6$Li, $\ket{g_1}$ is the $^1\Sigma^+_g\,(N=0)$ $v=38$, vibrational state, which is responsible for the Feshbach resonance. Starting from that state,  the best Franck-Condon factor~\cite{CoteJMS1999} arises for an optical transition to the  excited $A^1\Sigma^+_u\, (N=1)$ $v'=68$ vibrational state, which we take as $\ket{e}$. For the two-field experiments, we take  $\ket{g_2}$ to be the $v=37$ vibrational state, which is essentially uncoupled to the triplet state, as it lies $\simeq 57$ GHz below the resonant state. For the $v=38\rightarrow v'=68$ transition, the predicted oscillator strength is $f_{eg}=0.034$~\cite{CoteJMS1999}. We find that the corresponding Rabi frequency is $\Omega_1=2 \pi \times 5.6\,\text{MHz}\,\sqrt{I}$, where I is the intensity of the optical beam given in mW/mm$^2$.

We measure the atom number, the temperature, the diameter and power of the $\omega_1$ beam. For spatially uniform illumination, we use an $\omega_1$ beam with a $1/e^2$ intensity radius  of $750\,\mu$m, large compared to the size of the trapped cloud ($65-75\,\mu$m). Using the measured values, we fit {\it all} of the loss suppression data shown in Fig.~3~[a,b,c] of the main text using $\Omega_1=2 \pi \times c_1\,\text{MHz}\,\sqrt{I}$, with $c_1=5.45(0.10)$. The measured Rabi frequency is in good agreement with the value $5.6$ predicted using the Franck-Condon factors, based on the vibrational wave functions obtained from the molecular potentials~\cite{CoteJMS1999}. For Fig.~2~[a,b,c] and Fig.~4 of the main text, we find a smaller value $c_1=5.1$, most likely due to imperfect overlap of the $\Omega_1$ beam with the atom cloud.

Using $\Omega_1=2 \pi \times 5.1\,\text{MHz}\,\sqrt{I}$,  we see from Fig.~2~[a,b,c] of the main text that the continuum-dressed state model correctly predicts all three shifts of the narrow resonance, which is the dominant factor in determining $c_1$. Further,  for all three spectra, the amplitude and widths for both the broad and narrow resonances are correctly predicted without further adjustment. As the shift depends on $\Omega_1^2$ while the magnitudes of the loss rates depend on the product $\Omega_1^2\,|\langle g_1|E_k\rangle|^2$, the good agreement in the shape of the spectra confirms the transition strengths and momentum-dependent shifts predicted by the continuum-dressed state model  over the range from 540 G to 840 G.

For the loss-suppression measurements, we find the additional Rabi frequency for the $\omega_2$ beam, $\Omega_2=2 \pi \times c_2\,\text{MHz}\,\sqrt{I}$. Here, the  $1/e^2$ radius of the $\omega_2$ beam is $160\,\mu$m along the $z$-axis and $70\,\mu$m along the x-axis. Using the data of Fig.~[3] of the main text,  we find $c_2 = 0.14$, in reasonable agreement with the predicted value of $0.17$ for the very weak $v=37\rightarrow v'=68$ transition~\cite{Cotethesis}.

\end {document}